\documentclass[aps,pra,twocolumn,showpacs,superscriptaddress,draft,floatfix]{revtex4}
\input{epsf}
\usepackage{amsmath}
\usepackage{amssymb}
\input{epsf}

\newcommand{\rem}[1]{}

\begin{document}

\title{A simple representation of quantum process tomography}
\author{Giuliano Benenti}
\email{giuliano.benenti@uninsubria.it}
\affiliation{CNISM, CNR-INFM, and Center for Nonlinear and Complex Systems,
Universit\`a degli Studi dell'Insubria, via Valleggio 11, 22100 Como, Italy}
\affiliation{Istituto Nazionale di Fisica Nucleare, Sezione di Milano,
via Celoria 16, 20133 Milano, Italy}
\author{Giuliano Strini}
\email{giuliano.strini@mi.infn.it}
\affiliation{Dipartimento di Fisica, Universit\`a degli Studi di Milano,
via Celoria 16, 20133 Milano, Italy}
\date{\today}
\begin{abstract}
We show that the Fano representation leads to a particularly simple 
and appealing form of the quantum process tomography matrix $\chi_{_F}$,
in that the matrix $\chi_{_F}$ is real, the number of matrix 
elements is 
exactly equal to the number of free parameters required for the complete
characterization of a quantum operation, and these matrix elements
are directly related to evolution of the expectation values 
of the system's polarization measurements.
These facts are illustrated in
the examples of one- and two-qubit quantum noise channels.
\end{abstract}



\maketitle

\section{introduction}

The characterization of physical, generally noisy processes 
in open quantum systems is a key issue in quantum information 
science~\cite{qcbook,nielsen}. Quantum process tomography (QPT)
provides, in principle, full information on the dynamics of 
a quantum system and can be used to improve the design and
control of quantum hardware. 
Several QPT methods have been developed,
including the standard QPT~\cite{nielsen,chuangQPT,poyatos,korotkov},
ancilla-assisted QPT~\cite{dariano,leung,altepeter}, and direct 
characterization of quantum dynamics~\cite{lidar}.
In recent years QPT has been experimentally demonstrated 
with up to three-qubit systems 
in a variety of different implementations, including 
quantum optics~\cite{altepeter,mitchell,demartini,obrien,nambu,langford,kiesel,wang}, 
nuclear magnetic resonance quantum processors~\cite{childs,boulant,weinstein},
atoms in optical lattices~\cite{myrskog},
trapped ions~\cite{riebe,monz},
and solid-state qubits~\cite{howard,katz}.

Any quantum state $\rho$ can be expressed in the 
Fano form~\cite{fano,eberly,mahler}
(also known as Bloch representation). Since the density operator $\rho$
is Hermitian, the parameters of the expansion over the Fano basis are 
real. Furthermore, due to the linearity of quantum mechanics, any
quantum operation $\rho\to \rho'=\mathcal{E}(\rho)$ is represented,
in the Fano basis, by an affine map. 

In this paper, we point out that in standard QPT it is convenient to compute
the QPT matrix in the Fano basis. Such process matrix, $\chi_{_F}$, has the 
following advantages: (i) the matrix elements of $\chi_{_F}$ are real and
(ii) the number of matrix elements in $\chi_{_F}$ is exactly equal to
the number of free parameters needed in order to determine a generic 
quantum operation. 
Furthermore, the $\chi_{_F}$-matrix elements are directly
related to the modification, induced by the quantum operation 
$\mathcal{E}$, of the expectation values of the system's 
polarization measurements.
We will illustrate our results in the examples of 
one- and two-qubit quantum noise. In particular, we will determine 
in the $\chi_{_F}$-matrix the specific patterns of various quantum noise 
processes. Finally, we will discuss the number of free parameters 
physically relevant in determining a quantum operation for a 
two-qubit system exposed to weak local noise.

\section{Fano rapresentation of the standard QPT}
 
To simplify writing, we discuss the Fano representation
of the standard QPT only for qubits, even though the obtained results
can be readily extended to qudit systems. 
Any $n$-qubit state $\rho$  can be written in the 
Fano form as follows~\cite{fano,eberly,mahler}:
\begin{equation}
\rho=\frac{1}{N} \sum_{\alpha_1,...,\alpha_n=x,y,z,I}
c_{\alpha_1...\alpha_n}
\sigma_{\alpha_1}\otimes
\cdots \otimes \sigma_{\alpha_n},
\label{eq:fanoform}
\end{equation}
where 
$N=2^n$,
$\sigma_x$, $\sigma_y$, and $\sigma_z$ are the Pauli matrices, 
$\sigma_I\equiv \openone$, and 
\begin{equation}
c_{\alpha_1...\alpha_n}=
{\rm Tr}(\sigma_{\alpha_1}\otimes 
\cdots \otimes \sigma_{\alpha_n} \rho).
\end{equation}
Note that the normalization condition ${\rm Tr}(\rho)=1$ implies 
$c_{I...I}=1$. Moreover, the generalized 
Bloch vector ${\bf b}=\{b_\alpha\}_{\alpha=1,...,N^2-1}$
is real due to the hermiticity of $\rho$. 
Here $b_\alpha\equiv c_{\alpha_1...\alpha_n}$, 
with $\alpha \equiv \sum_{k=1}^{n} i_k 4^{n-k}$,
where we have defined $i_k=1,2,3,4$ in correspondence
to $\alpha_k=x,y,z,I$. 
Note that from $1$ to $n$ qubits run from the most significant
to the least significant.
For instance, for two qubits ($n=2$),
the $N^2-1=15$ components of vector ${\bf b}$ are ordered as
follows: 
\begin{equation}
\begin{array}{c}
{\bf b}^T=(b_1,b_2,...,b_{15})=
(c_{xx},
c_{xy},
c_{xz},
c_{xI},
c_{yx},
c_{yy},
\\
c_{yz},
c_{yI},
c_{zx},
c_{zy},
c_{zz},
c_{zI},
c_{Ix},
c_{Iy},
c_{Iz}).
\end{array}
\end{equation}

Due to the linearity of quantum mechanics any quantum operation
$\rho\to \rho'=\mathcal{E}(\rho)$ 
is represented in the Fano basis 
$\{\sigma_{\alpha_1}\otimes ...\otimes\sigma_{\alpha_n}\}$ 
by an affine map:
\begin{equation}
\left[
\begin{array}{c}
{\bf b'}
\\
\hline
1
\end{array}
\right]
=
\mathcal{M} 
\left[
\begin{array}{c}
{\bf b}
\\
\hline
1
\end{array}
\right]
=
\left[
\begin{array}{ccc}
  {\bf M} & \Big \lvert & {\bf a}  \\
 \hline 
  {\bf 0}^T & \Big \lvert & 1 
\end{array}
\right]
\left[
\begin{array}{c}
{\bf b}
\\
\hline
1
\end{array}
\right],
\end{equation} 
where ${\bf M}$ is a $(N^2-1) \times (N^2-1)$ matrix,
${\bf a}$ a column vector of dimension $N^2-1$ and 
${\bf 0}$ the null vector of the same dimension.

All information about the quantum operation  
$\mathcal{E}$ is contained in the $N^4-N^2$ free elements 
of matrix $\mathcal{M}$, namely in the matrix 
\begin{equation}
\chi_{_F}=
\left[
\begin{array}{ccc}
  {\bf M} & \Big \lvert & {\bf a}  
\end{array}
\right].
\label{eq:processmatrix}
\end{equation}
To obtain the QPT matrix $\chi_{_F}$ from experimental data, one
needs to prepare $N^2$ linearly independent initial states
$\{\rho_i\}$, let
them evolve according to the quantum operation $\mathcal{E}$ and
then measure the resulting states
$\{\rho_i'=\mathcal{E}(\rho_i)\}$. 
If we call $\mathcal{R}$ the $N^2\times N^2$
matrix whose columns are given by the  
Fano representation of states $\rho_i$
and $\mathcal{R'}$ the corresponding matrix constructed from states 
$\rho_i'$, we have
\begin{equation}
\mathcal{R'}=\mathcal{M} \mathcal{R},
\end{equation}
and therefore 
\begin{equation}
\mathcal{M}=\mathcal{R'} \mathcal{R}^{-1}.
\end{equation}

As it is well known~\cite{nielsen}, the standard QPT can be performed 
with initial states being product states and local measurements 
of the final states. As initial states $\{\rho_i\}$ 
we choose the $4^n$ tensor-product states 
of the $4$ single-qubit states 
\begin{equation}
|0\rangle, \;\;\;|1\rangle, \;\;\;\frac{1}{\sqrt{2}}(|0\rangle +|1\rangle),
\;\;\;\frac{1}{\sqrt{2}}(|0\rangle +i |1\rangle).
\label{sepbasis}
\end{equation}
To estimate $\mathcal{R'}$, one needs to prepare many copies of each initial
state $\rho_i$, let them evolve according to the quantum 
operation $\mathcal{E}$ and then measure observables 
$\sigma_{\alpha_1}\otimes \cdots \otimes \sigma_{\alpha_n}$.
Of course, such measurements can be performed on the computational
basis $\{|0\rangle,|1\rangle\}^{\otimes n}$, provided 
each measurement is preceded by suitable single-qubit rotations.

\section{Single-qubit systems}

The matrix $\mathcal{R}$ corresponding to basis (\ref{sepbasis})
reads
\begin{equation}
\mathcal{R}=\left[
\begin{array}{cccc}
 0 & 0  & 1 & 0 \\
 0 & 0  & 0 & 1 \\
 1 & -1 & 0 & 0 \\
 1 & 1  & 1 & 1
\end{array}
\right].
\end{equation}
Therefore,
\begin{equation}
\mathcal{R}^{-1}=\left[
\begin{array}{cccc}
 -\frac{1}{2} & -\frac{1}{2}  & \frac{1}{2}  & \frac{1}{2} \\
 -\frac{1}{2} & -\frac{1}{2}  & -\frac{1}{2} & \frac{1}{2} \\
 1            & 0             & 0            & 0 \\
 0            & 1             & 0            & 0 
\end{array}
\right].
\end{equation}

The coefficients $(c_x,c_y,c_z)$ in the Fano form 
(\ref{eq:fanoform}) are the Bloch-vector coordinates 
of the density matrix $\rho$ in the Bloch-ball representation of 
single-qubit states.
We need $N^4-N^2=12$ parameters to characterize a generic quantum operation
acting on a single qubit. Each parameter describes a particular noise 
channel (like bit flip, phase flip, amplitude damping,...) and can be 
most conveniently visualized as associated with rotations, deformations 
and displacements of the Bloch ball~\cite{qcbook,nielsen,BFS}.
Here we point out that these noise channels 
lead to specific patters in the state process matrix 
$\chi_{_F}$.

For instance, for the phase-flip channel,
\begin{equation}
\rho'=\mathcal{E}(\rho)=
p \sigma_z \rho \sigma_z + (1-p) \rho,
\;\;
(0\le p \le \frac{1}{2}),
\label{eq:phaseflip}
\end{equation}
we have 
\begin{equation}
\mathcal{R'}=\left[
\begin{array}{cccc}
 0 & 0  & 1-2p & 0 \\
 0 & 0  & 0 & 1-2p \\
 1 & -1 & 0 & 0 \\
 1 & 1  & 1 & 1
\end{array}
\right].
\end{equation}
We can then compute $\mathcal{M}=\mathcal{R'}R^{-1}$, and the first 
three lines of $\mathcal{M}$ correspond to the state matrix
\begin{equation}
\chi_{_F}^{(\rm pf)}=
\left[
\begin{array}{cccc}
 1-2p & 0 & 0 & 0 \\
 0 & 1-2p & 0 & 0 \\
 0 & 0 & 1 & 0
\end{array}
\right].
\end{equation}
Therefore, the Bloch ball is mapped into an ellipsoid 
with $z$ as symmetry axis: 
\begin{equation}
\left\{
\begin{array}{l}
c_x\to c_x'=(1-2p)c_x,\\
c_y\to c_y'=(1-2p)c_y,\\
c_z\to c_z'=c_z.
\end{array}
\right.
\end{equation}

As a further example, we consider the amplitude damping channel:
\begin{equation}
\rho'=\sum_{k=0}^1 E_k \rho E_k^\dagger, 
\end{equation}
with the Kraus operators
\begin{equation}
E_0=|0\rangle\langle 0| +\sqrt{1-p} |1\rangle\langle 1|,
\;
E_1=\sqrt{p} |0\rangle\langle 1|,
\;
(0\le p \le 1).
\end{equation}
In this case we obtain 
\begin{equation}
\chi_{_F}^{(\rm ad)}=
\left[
\begin{array}{cccc}
 \sqrt{1-p} & 0 & 0 & 0 \\
 0 & \sqrt{1-p} & 0 & 0 \\
 0 & 0 & 1-p & p 
\end{array}
\right].
\end{equation}
The Bloch ball is deformed into an ellipsoid, with its center
displaced along the $z$-axis: 
\begin{equation}
\left\{
\begin{array}{l}
c_x\to c_x'=\sqrt{1-p}c_x,\\
c_y\to c_y'=\sqrt{1-p}c_y,\\
c_z\to c_z'=(1-p)c_z+p.
\end{array}
\right.
\end{equation}

\section{Two-qubit systems}

\begin{widetext}
Matrices $\mathcal{R}$ and $\mathcal{R}^{-1}$ 
corresponding to the $16$ tensor-product states of 
single-qubit states (\ref{sepbasis}) read as follows:
\begin{equation}
\mathcal{R}=
\left [
\begin{array}{cccccccccccccccc}
0  &  0  &  0  &  0  &  0  &  0  &  0  &  0  & 
0  &  0  &  1  &  0  &  0  &  0  &  0  &  0  \\
0  &  0  &  0  &  0  &  0  &  0  &  0  &  0  & 
0  &  0  &  0  &  1  &  0  &  0  &  0  &  0  \\
0  &  0  &  0  &  0  &  0  &  0  &  0  &  0  & 
1  &  -1 &  0  &  0  &  0  &  0  &  0  &  0  \\
0  &  0  &  0  &  0  &  0  &  0  &  0  &  0  & 
1  &  1  &  1  &  1  &  0  &  0  &  0  &  0  \\
0  &  0  &  0  &  0  &  0  &  0  &  0  &  0  & 
0  &  0  &  0  &  0  &  0  &  0  &  1  &  0  \\
0  &  0  &  0  &  0  &  0  &  0  &  0  &  0  & 
0  &  0  &  0  &  0  &  0  &  0  &  0  &  1  \\
0  &  0  &  0  &  0  &  0  &  0  &  0  &  0  & 
0  &  0  &  0  &  0  &  1  &  -1 &  0  &  0  \\
0  &  0  &  0  &  0  &  0  &  0  &  0  &  0  & 
0  &  0  &  0  &  0  &  1  &  1  &  1  &  1  \\
0  &  0  &  1  &  0  &  0  &  0  &  -1 &  0  & 
0  &  0  &  0  &  0  &  0  &  0  &  0  &  0  \\
0  &  0  &  0  &  1  &  0  &  0  &  0  &  -1 & 
0  &  0  &  0  &  0  &  0  &  0  &  0  &  0  \\
1  &  -1 &  0  &  0  &  -1 &  1  &  0  &  0  & 
0  &  0  &  0  &  0  &  0  &  0  &  0  &  0  \\
1  &  1  &  1  &  1  &  -1 &  -1 &  -1 &  -1 & 
0  &  0  &  0  &  0  &  0  &  0  &  0  &  0  \\
0  &  0  &  1  &  0  &  0  &  0  &  1  &  0  & 
0  &  0  &  1  &  0  &  0  &  0  &  1  &  0  \\
0  &  0  &  0  &  1  &  0  &  0  &  0  &  1  & 
0  &  0  &  0  &  1  &  0  &  0  &  0  &  1  \\
1  &  -1 &  0  &  0  &  1  &  -1 &  0  &  0  & 
1  &  -1 &  0  &  0  &  1  &  -1 &  0  &  0  \\
1  &  1  &  1  &  1  &  1  &  1  &  1  &  1  & 
1  &  1  &  1  &  1  &  1  &  1  &  1  &  1  
\end{array}
\right],
\end{equation}
\begin{equation}
\mathcal{R}^{-1}=\frac{1}{4}
\left [
\begin{array}{cccccccccccccccc}
1  &  1  &  -1 &  -1 &  1  &  1  &  -1 &  -1 & 
-1 &  -1 &  1  &  1  &  -1 &  -1 &  1  &  1  \\
1  &  1  &  1  &  -1 &  1  &  1  &  1  &  -1 & 
-1 &  -1 &  -1 &  1  &  -1 &  -1 &  -1 &  1  \\
-2 &  0  &  0  &  0  &  -2 &  0  &  0  &  0  & 
2  &  0  &  0  &  0  &  2  &  0  &  0  &  0  \\
0  &  -2 &  0  &  0  &  0  &  -2 &  0  &  0  & 
0  &  2  &  0  &  0  &  0  &  2  &  0  &  0  \\
1  &  1  &  -1 &  -1 &  1  &  1  &  -1 &  -1 & 
1  &  1  &  -1 &  -1 &  -1 &  -1 &  1  &  1  \\
1  &  1  &  1  &  -1 &  1  &  1  &  1  &  -1 & 
1  &  1  &  1  &  -1 &  -1 &  -1 &  -1 &  1  \\
-2 &  0  &  0  &  0  &  -2 &  0  &  0  &  0  & 
-2 &  0  &  0  &  0  &  2  &  0  &  0  &  0  \\
0  &  -2 &  0  &  0  &  0  &  -2 &  0  &  0  & 
0  &  -2 &  0  &  0  &  0  &  2  &  0  &  0  \\
-2 &  -2 &  2  &  2  &  0  &  0  &  0  &  0  & 
0  &  0  &  0  &  0  &  0  &  0  &  0  &  0  \\
-2 &  -2 &  -2 &  2  &  0  &  0  &  0  &  0  & 
0  &  0  &  0  &  0  &  0  &  0  &  0  &  0  \\
4  &  0  &  0  &  0  &  0  &  0  &  0  &  0  & 
0  &  0  &  0  &  0  &  0  &  0  &  0  &  0  \\
0  &  4  &  0  &  0  &  0  &  0  &  0  &  0  & 
0  &  0  &  0  &  0  &  0  &  0  &  0  &  0  \\
0  &  0  &  0  &  0  &  -2 &  -2 &  2  &  2  & 
0  &  0  &  0  &  0  &  0  &  0  &  0  &  0  \\
0  &  0  &  0  &  0  &  -2 &  -2 &  -2 &  2  & 
0  &  0  &  0  &  0  &  0  &  0  &  0  &  0  \\
0  &  0  &  0  &  0  &  4  &  0  &  0  &  0  & 
0  &  0  &  0  &  0  &  0  &  0  &  0  &  0  \\
0  &  0  &  0  &  0  &  0  &  4  &  0  &  0  & 
0  &  0  &  0  &  0  &  0  &  0  &  0  &  0  
\end{array}
\right].
\end{equation}
\end{widetext}

The coordinates $\{c_{\alpha_1\alpha_2}\}$ in the Fano form (\ref{eq:fanoform})
are the expectation values of the polarization measurements 
$\{\sigma_{\alpha_1}\otimes \sigma_{\alpha_2}\}$. The coefficients in the 
state matrix $\chi_{_F}$ representing a quantum operation $\mathcal{E}$
can therefore be interpreted in terms of modification of these expectation 
values. 

For instance, let us assume that the two qubits are 
independently exposed to pure dephasing, that is,
to quantum noise described 
by the phase-flip channel (\ref{eq:phaseflip}), 
with the same noise strength 
$p$ for both qubits. 
The process matrix for such uncorrelated dephasing channel is
given by
\begin{equation}
\chi_{_F}^{({\rm ud})}=
\left [
\begin{array}{ccccccccccccccc}
g^2  &   0   &   0   &   0   &   0   &   0   &   0   &   0   & 
0    &   0   &   0   &   0   &   0   &   0   &   0   \\
0    &   g^2 &   0   &   0   &   0   &   0   &   0   &   0   & 
0    &   0   &   0   &   0   &   0   &   0   &   0   \\
0    &   0   &   g   &   0   &   0   &   0   &   0   &   0   & 
0    &   0   &   0   &   0   &   0   &   0   &   0   \\
0    &   0   &   0   &   g   &   0   &   0   &   0   &   0   & 
0    &   0   &   0   &   0   &   0   &   0   &   0   \\
0    &   0   &   0   &   0   &   g^2 &   0   &   0   &   0   & 
0    &   0   &   0   &   0   &   0   &   0   &   0   \\
0    &   0   &   0   &   0   &   0   &   g^2 &   0   &   0   & 
0    &   0   &   0   &   0   &   0   &   0   &   0   \\
0    &   0   &   0   &   0   &   0   &   0   &   g   &   0   & 
0    &   0   &   0   &   0   &   0   &   0   &   0   \\
0    &   0   &   0   &   0   &   0   &   0   &   0   &   g   & 
0    &   0   &   0   &   0   &   0   &   0   &   0   \\
0    &   0   &   0   &   0   &   0   &   0   &   0   &   0   & 
g    &   0   &   0   &   0   &   0   &   0   &   0   \\
0    &   0   &   0   &   0   &   0   &   0   &   0   &   0   & 
0    &   g   &   0   &   0   &   0   &   0   &   0   \\
0    &   0   &   0   &   0   &   0   &   0   &   0   &   0   & 
0    &   0   &   1   &   0   &   0   &   0   &   0   \\
0    &   0   &   0   &   0   &   0   &   0   &   0   &   0   & 
0    &   0   &   0   &   1   &   0   &   0   &   0   \\
0    &   0   &   0   &   0   &   0   &   0   &   0   &   0   & 
0    &   0   &   0   &   0   &   g   &   0   &   0   \\
0    &   0   &   0   &   0   &   0   &   0   &   0   &   0   & 
0    &   0   &   0   &   0   &   0   &   g   &   0   \\
0    &   0   &   0   &   0   &   0   &   0   &   0   &   0   & 
0    &   0   &   0   &   0   &   0   &   0   &   1   \\
\end{array}
\right],
\label{eq:chiud}
\end{equation}
where $g\equiv 1-2p$. 
Correspondingly, the mapping for the expectation values of the 
polarization measurements reads
\begin{equation}
c_{\alpha_1\alpha_2}'=g^{m_1+m_2} c_{\alpha_1\alpha_2},
\end{equation}
where $m_i=1$ for $\alpha_i=x,y$ and 
$m_i=0$ for $\alpha_i=z,I$.

As an example of nonlocal quantum noise, we consider a model 
of fully correlated pure dephasing. We model the interaction of the 
two qubits with the environment as a phase-kick rotating both 
qubits through the same angle $\theta$ about the $z$ axis of 
the Bloch ball. This rotation is described in the 
$\{|0\rangle,|1\rangle\}$ basis by the unitary matrix
\begin{equation}
R_z(\theta)=
\left[
\begin{array}{cc}
e^{-i\frac{\theta}{2}} & 0 \\
0 & e^{i\frac{\theta}{2}}
\end{array}
\right]
\otimes
\left[
\begin{array}{cc}
e^{-i\frac{\theta}{2}} & 0 \\
0 & e^{i\frac{\theta}{2}}
\end{array}
\right].
\end{equation}
We assume that the rotation angle is drawn from the random distribution
\begin{equation}
p(\theta)=\frac{1}{\sqrt{4\pi\lambda}} e^{-\frac{\theta^2}{4\lambda}}.
\end{equation}
Therefore, the final state $\rho'$, obtained after averaging
over $\theta$, is given by
\begin{equation}
\rho'=\int_{-\infty}^{+\infty}d\theta p(\theta)
R_z(\theta) \rho R_z^\dagger(\theta).
\end{equation} 
For this correlated dephasing channel we obtain 
the process matrix 
\begin{equation}
\chi_{_F}^{({\rm cd})}=
\left [
\begin{array}{ccccccccccccccc}
h    &   0   &   0   &   0   &   0   &   k   &   0   &   0   & 
0    &   0   &   0   &   0   &   0   &   0   &   0   \\
0    &   h   &   0   &   0   &   -k  &   0   &   0   &   0   & 
0    &   0   &   0   &   0   &   0   &   0   &   0   \\
0    &   0   &   g   &   0   &   0   &   0   &   0   &   0   & 
0    &   0   &   0   &   0   &   0   &   0   &   0   \\
0    &   0   &   0   &   g   &   0   &   0   &   0   &   0   & 
0    &   0   &   0   &   0   &   0   &   0   &   0   \\
0    &   -k  &   0   &   0   &   h   &   0   &   0   &   0   & 
0    &   0   &   0   &   0   &   0   &   0   &   0   \\
k   &   0   &   0   &   0   &   0   &   h   &   0   &   0   & 
0    &   0   &   0   &   0   &   0   &   0   &   0   \\
0    &   0   &   0   &   0   &   0   &   0   &   g   &   0   & 
0    &   0   &   0   &   0   &   0   &   0   &   0   \\
0    &   0   &   0   &   0   &   0   &   0   &   0   &   g   & 
0    &   0   &   0   &   0   &   0   &   0   &   0   \\
0    &   0   &   0   &   0   &   0   &   0   &   0   &   0   & 
g    &   0   &   0   &   0   &   0   &   0   &   0   \\
0    &   0   &   0   &   0   &   0   &   0   &   0   &   0   & 
0    &   g   &   0   &   0   &   0   &   0   &   0   \\
0    &   0   &   0   &   0   &   0   &   0   &   0   &   0   & 
0    &   0   &   1   &   0   &   0   &   0   &   0   \\
0    &   0   &   0   &   0   &   0   &   0   &   0   &   0   & 
0    &   0   &   0   &   1   &   0   &   0   &   0   \\
0    &   0   &   0   &   0   &   0   &   0   &   0   &   0   & 
0    &   0   &   0   &   0   &   g   &   0   &   0   \\
0    &   0   &   0   &   0   &   0   &   0   &   0   &   0   & 
0    &   0   &   0   &   0   &   0   &   g   &   0   \\
0    &   0   &   0   &   0   &   0   &   0   &   0   &   0   & 
0    &   0   &   0   &   0   &   0   &   0   &   1   \\
\end{array}
\right],
\label{eq:chicd}
\end{equation}
where $g\equiv e^{-\lambda}$, 
$h\equiv \frac{1}{2}(1+g^4)$, $k\equiv\frac{1}{2}(1-g^4)$.
It is clear that the process matrix (\ref{eq:chicd})
for correlated dephasing
has a pattern that allows to clearly distinguish it from
the process matrix (\ref{eq:chiud}) 
for the uncorrelated dephasing.

It is also obvious that, if there exists partial previous knowledge
of the dominant noise sources, it is not necessary to construct
the whole state process matrix $\chi_{_F}$ in order to characterize
the quantum operation. For instance, if we know a priori that 
dephasing is the main source of noise and we wish to estimate its
degree of correlation, it is sufficient to prepare, for instance, the
initial state $\rho=\frac{1}{2}(|0\rangle+|1\rangle)^{\otimes 2}$ 
and measure the $x$- and $y$-polarizations of both qubits for 
the final state $\rho'$. The initial state is fully polarized along $x$,
and therefore
\begin{equation}
\left\{
\begin{array}{l}
c_{xx}=1, \\ 
c_{yy}=0.
\end{array}
\right.
\end{equation}
For the final state, in the case of fully correlated dephasing 
\begin{equation}
\left\{
\begin{array}{l}
(c_{xx}')^{({\rm cd})}=h=\frac{1}{2}[1+g^4],\\
(c_{yy}')^{({\rm cd})}=k=\frac{1}{2}[1-g^4],
\end{array}
\right.
\end{equation}
while the expectation values of the $xx$- and $yy$-polarization 
measurements are remarkably different for uncorrelated dephasing:
\begin{equation}
\left\{
\begin{array}{l}
(c_{xx}')^{({\rm ud})}=g^2,\\
(c_{yy}')^{({\rm ud})}=0.
\end{array}
\right.
\end{equation}

While in general two-qubit quantum operations depend on 
$N^4-N^2=240$ real parameters, an important question is how
many parameters are physically significant. The answer of 
course depends on the specific noise processes. However, a
clear answer can be given assuming that external noise is 
weak and local, that is to say, it acts independently on 
the two qubits. In this case, local noise is described by $24$
parameters, $12$ for each qubit. Undesired coupling effects
(cross talk) between qubits can be characterized with only
three additional 
parameters, $\theta_x$, $\theta_y$, and $\theta_z$. 
Indeed, any two-qubit unitary transformation $U$ 
can be decomposed as~\cite{khaneja,kraus,nielsenPRA}
\begin{equation}
U=(A_1\otimes B_1) e^{i(\theta_x \sigma_x\otimes \sigma_x+
\theta_y \sigma_y\otimes \sigma_y +\theta_z \sigma_z\otimes \sigma_z)}
(A_2\otimes B_2),
\end{equation}
with $A_1$, $A_2$, $B_1$, and $B_2$ appropriate single-qubit
unitaries. 
In the limit of weak noise, the state matrix $\chi_{_F}$
is simply given by the sum of the contributions of each noise 
channel. Therefore, the local unitaries 
$A_1$, $A_2$, $B_1$, and $B_2$ only change the $24$ local noise 
parameters and overall we need $24+3=27\ll 240$ parameters 
to describe the quantum noise. In the symmetric case in which
the local noise parameters are the same for both qubits the
number of free parameters further reduces to $12+3=15$. 
The above argument can be easily extended to many-qubit 
systems.
Due to the two-body nature of interactions, we need to
determine ${\mathcal N}=12n+3\frac{n(n-1)}{2}$ parameters 
to characterize noise. Note that 
${\mathcal N}=O(\log N)\ll N^4-N^2$.
Of course, cases with strong or nonlocal noise 
would require a larger number of free parameters.

\section{Conclusions}

We have shown that the Fano representation of the standard QPT
is convenient, since the process matrix $\chi_{_F}$ is real 
and the number of matrix elements is exactly equal to the number
of free parameters required for the complete characterization
of a generic quantum operation. 
Moreover, the matrix elements of $\chi_{_F}$ are directly related to 
the evolution, induced by the quantum operation, of the 
system's polarization measurements. 
We have also shown that quantum noise channels have specific 
patterns in the Fano representation of $\chi_{_F}$. 
Finally, we have shown that in the case, of interest
for quantum information processing, of weak and local noise
the number of relevant noise parameters is 
$\mathcal{N}=O(\log N)\ll N^4-N^2$, that is, much smaller 
than the number of parameters needed to determine a 
generic quantum operation. In this case, 
the $\chi_{_F}$-matrix is very sparse and therefore the 
number of polarization measurements needed to reconstruct it
is much smaller than for a generic quantum operation, thus
considerably reducing the QPT complexity.

\end{document}